\documentclass[%
 reprint,
 amsmath,amssymb,
 aps,
prb,
]{revtex4-1}

\usepackage{graphicx}
\usepackage{dcolumn}
\usepackage{bm}

\begin{document}

\preprint{APS/123-QED}

\setlength{\abovecaptionskip}{-60pt}

\title{An Effective Spin Hamiltonian Approach to Metamagnetism -I}

\author{P. Kumar}

\affiliation{ Department of Physics, University of Florida, Gainesville, FL. 32611-8440}

\author{B. S. Shivaram and V. Celli}

\affiliation{ Department of Physics, University of Virginia, Charlottesville, VA. 22904 }

\date{\today}

\begin{abstract}

We describe a minimal model, based on a spin only Hamiltonian with a single energy scale for itinerant electron metamagnetism.  Within this model the metamagnetic critical field is directly proportional to the temperature where a peak in the linear susceptibility occurs which in turn is related in a simple manner to the temperature where the nonlinear susceptibilities also peak.  The spin dependent thermodynamic properties are derived in a straightforward manner and bear a striking resemblance to observations in such strongly correlated systems as heavy fermion materials. We also consider extensions of the model by including effects such as a mean field to encompass observed deviations from a minimal metamagnetic behavior.

\begin{description}

\item[PACS numbers]

75.30.Mb, 75.20.Hr

\end{description}

\end{abstract}

\pacs{Valid PACS appear here}

\maketitle

\section {Introduction}
Metamagnetism refers to the response of a system to an external magnetic field where the field dependent magnetization at low enough temperature suddenly increases from a small value to a large value at a critical field.  In many materials the transition is sharp at temperature $T = 0$ but generally gets smoother with increasing temperature.  It is a quantum phase transition which takes place at ${B = B_c}$ and $T = 0$. At any finite temperature, there is no metamagnetic transition but only a point of inflection in the magnetization which disappears at higher temperatures. \\
\indent
The phenomenon of metamagnetism is seen in many diverse solid state systems \cite {GiordanoRev1977}.  In the (insulating) transition metal compounds such as $\mathrm {FeCl_2}$ and $\mathrm{DyPO_4}$ it is ascribed to an anisotropic antiferromagnetic ground state where the antiferromagnetic correlations are destroyed by an external magnetic field.  The correlations arise from inter-ionic, usually antiferromagnetic (AFM) exchange interaction while the anisotropy comes from the crystalline electric fields.   When the magnetic field is applied in a specific direction, the spins respond to it only after the field increases beyond a certain threshold, which may depend on the direction.  Since the tendency for the spins to resist alignment are governed by the exchange interaction, the critical magnetic field should scale with the AFM ordering temperature.  The effective Hamiltonian is a spin Hamiltonian based on real spins and the crystal electric fields are dependent on the real lattice structure.\\
\indent
Metamagnetic properties in strongly correlated electronic materials such as heavy fermion systems (HFS) have a different manifestation \cite{AokiRev2013}.  Here the principal energy scale is the ion-electron “Kondo” interaction and the principal correlation must be something related to Kondo temperature. Kondo compensation for a single impurity spin is a crossover effect, i.e., the observables tend to change smoothly with temperature and field.  There is no phase transition. However, there could be a (quantum) field induced phase transition for conduction electrons in a “lattice” of spins as is the case for HFS.  The critical field $B_c [\mathrm{Tesla}]$ in such compounds has been shown by Hirose et al. \cite{Hirose2011, Saito2000} to scale with $ \approx 1.5\  T_1[\mathrm{Kelvin}]$ where $T_1$ is the temperature where a peak in the linear susceptibility is invariably present.  The critical field does not seem to depend on the lattice structure or other microscopic details of the system. Hirose et al. have also shown the inverse of the peak value in the susceptibility also scales in a manner similar to that found earlier in intermetallic compounds \cite {AndersonAdvPhys1997}.  \\
\indent
In more recent work\cite {ShivaramChi5PRB2014, ShivaramUniversalityPRB2014} Shivaram et al.\ introduced a comprehensive study of the nonlinear magnetic field response in HFS.  In a paramagnet, the third order susceptibility is negative; in a metamagnet, one might expect it to become positive at temperatures below $T_1$. Indeed,  Shivaram et al.\ found it has a peak at a temperature $T_3$ which scales with $T_1$,  i.e., $T_3 \approx 0.5\, T_1$.  Thus there appears to be a single energy scale involved that governs the behavior of the susceptibilities.  The earliest theoretical suggestion of a single energy scale in the context of strongly correlated electronic systems may be found in the scaling suggestions of Wohlleben et al  \cite {ZieglowskiPRL1986} and Thalmeier and Fulde \cite {ThalmeierFuldeEPL1986} .  There have also been other scaling approaches \cite {ContinentinoPRB1989, ContinentinoJPhysique1991}.  \\
 \indent 
In a follow up publication, Shivaram et al.\ have also studied the sound velocity in UPt$_3$ as a function of magnetic field and found anomalies in the vicinity of the critical field \cite {ShivaramPRB2015}.  The principal conclusion from all these recent experiments is that, to a good approximation, there is a single energy scale and therefore a simple, minimal effective Hamiltonian.\\  
 \indent 
The purpose of this paper is to provide in a comprehensive manner the results that may be obtainable from a minimal model, where metamagnetism is viewed as resulting from a crossing of two energy levels, one with a small magnetic moment and another with a large one, at the critical field.  Specifically, we consider an effective spin S = 1 model whose energy levels are separated by $\Delta$ and an effective Hamiltonian 

\begin{equation} 
 H=\Delta  S_{z}^2 -  \gamma B S_{z}    \label{eq:Hz}
\end{equation} 

\noindent with $\Delta>0$.  Here metamagnetism happens at $\mathrm{B_c} = \Delta/\gamma$.  It is sharp at T = 0 but broadens as the temperature increases.  As will be shown in Sect. II, the temperature dependence of the linear susceptibility has a peak at $\mathrm{T_1}=2\Delta/3k$.  The nonlinear susceptibilities, generally expected to be small, turn positive as the temperature decreases below $\mathrm{T_1}$ and have a sizeable peak at temperatures similar to those seen in experiments \cite{ParkJPCM1994, BauerPRB2006}.  The linear susceptibility, in particular its value at the peak, is inversely proportional to $\Delta$, same as the correlation noted by Saito et al.\cite {Saito2000}.  Likewise, the nonlinear susceptibilities should have peak values inversely proportional to $\Delta^n$ (where n = 3 or 5 for the third or fifth order susceptibility, respectively).  
It is understood that the effective spin Hamiltonian \ref{eq:Hz} is meant to describe metamagnetism in a highly simplified way, at temperatures not much greater than $\Delta/k$. The Hamiltonian is a standard anisotropic case, one that might arise from an interaction that breaks rotational symmetry of the spin space.  In a physical realization of this model there is a localized spin oriented in the z-direction dictated by a crystal field and an itinerant electron, s or p like, antiferromagnetically coupled to the localized spin.  The effects on pressure, or more generally on strains, are included through a deformation dependence of $\Delta$.  The z-axis is simply any direction along which metamagnetism occurs. A simple metamagnetic Hamiltonian is one that reduces to \ref{eq:Hz}, at least approximately, for one or more directions, possibly with different $\Delta$ and/or $\gamma$ values in different directions. \\

This paper is organized as follows:  Section II contains a discussion of the model we propose and the associated spin Hamiltonian.  The model is used to calculate the Helmholtz free energy which then leads to an evaluation of a number of thermodynamic observables as well as the nonlinear susceptibilities.  Section III presents extensions to the minimal model by including a mean field as well as by broadening the energy levels.  Where possible we try to provide references to experimental results mostly citing those observed in the heavy fermion metamagnets, UPt$_3$ and CeRu$_2$Si$_2$.  This is done for convenience but similar metamagnetic responses maybe found in a wide variety of strongly correlated d and f-electron itinerant systems.  Sec. IV contains a summary of our results and a discussion of the assumptions and contexts in which they were obtained.  An appendix is included in the end with additional results of interest.
\\

\section{The Minimal Model}
Consider a localized spin 1, resulting from the combination of two or more half-integer spins, that is forced by a crystal field to lie in the xy plane, i.e., away from the z-direction. There is no interaction with other spins in the material, which may be crystalline or amorphous.   
The Hamiltonian is

\begin{equation} 
\mathcal{H}=\Delta  S_{z}^2 -  \gamma\,\bm{B \cdot S}   \label{eq:Hmin}
\end{equation} 
This reduces to Eq.~\ref{eq:Hz} when $\bm{B}$ is in the z direction and represents the simplest example of metamagnetism. There is only one energy scale, $\Delta$. For $\Delta > 0$ the ground state is nonmagnetic but the excited state, separated by energy $\Delta$, is endowed with a full moment. As we show below, the metamagnetic phase transition occurs only in one direction, the z axis selected by the crystal field. 
 We consider this to be the "minimal model" and we use it as the starting point for further work. More complex energy level schemes (instead of the singlet-doublet here) give similar results. \\

We first work out the magnetic observables in this model, such as the magnetization  and the susceptibilities of various order.  We next evaluate other observables such as the heat capacity and bulk modulus. The (possible) dependence of $\Delta$ and $\gamma$ on the volume $V$ is touched upon in the end.\\

\subsection{Magnetization, susceptibilities}
We start with the simple cases when $\bm{B}$ is parallel and perpendicular to z. We next evaluate the more general case of an arbitrary field direction.
\subsubsection{B parallel to z}
To proceed we need the three energy eigenvalues $\epsilon_1 = 0, \epsilon_2 = \Delta -\gamma B$ and $\epsilon_3 = \Delta + \gamma B$. The non-magnetic level $\epsilon_1$ is the ground state until $B$ reaches the critical value $ B_c=\Delta/\gamma$, at which point the magnetic level $\epsilon_2$ becomes the ground state. This level crossing causes the metamagnetic phase transition.\\
The free energy $F(B,T)$ and the magnetization $M = - \partial F / \partial B$ are given by 

\begin{equation} 
F\left(B,T\right)=\ -kT{\ln \left[1+2e^{-\Delta /kT}{{\cosh} (\frac{\gamma B}{kT})\ }\right]\ } \label{eq:FforBz} 
\end{equation} 

\begin{equation} 
m=\ \frac{M}{\ \gamma }=\ \frac{\sinh\ \left(b/\tau \right)}{a + \cosh\left(b/\tau \right)}.  \label{eq:mforBz}
\end{equation} 
\noindent
Here and in the following $b = \gamma B / \Delta $ and $\tau = kT/ \Delta$; further, until new notice, $a=\frac{1}{2}e^{1/\tau }$.  As expected, at  low temperatures the magnetization rises rapidly around $b =1$, with 
the width of the rise gven by $2\tau$,  and there is a crossing of the magnetization isotherms at  $m=1/2$ (Fig.~1).  \\
The zero-field susceptibilities (at constant volume) also display the expected metamagnetic behavior. It is convenient to define dimensionless susceptibilities, linear as well as nonlinear, from the expansion $m = \chi_1 b + +\chi_3 b^3 +\chi_5 b^5$; the measured susceptibilities are $\gamma (\gamma/\Delta)^n \chi_{n}$.  In the present case one finds:

\begin{equation} 
{\chi }_{1z}=\ \frac{1}{\tau }\frac{1}{a+1}  \label{eq:chi1forBz}
\end{equation} 

\begin{equation} 
{\chi }_{3z}=\ \frac{1}{6\tau^3}\frac{a-2}{{\left(a+1\right)}^2}  \label{eq:chi3forBz}
\end{equation} 

\begin{equation} 
{\chi }_{5z}=\ \frac{1}{{120\tau }^5}\frac{a^2-13a+16}{{\left(a+1\right)}^3}  \label{eq:chi5forBz}
\end{equation} 

\begin{figure}
\includegraphics[width=110mm]{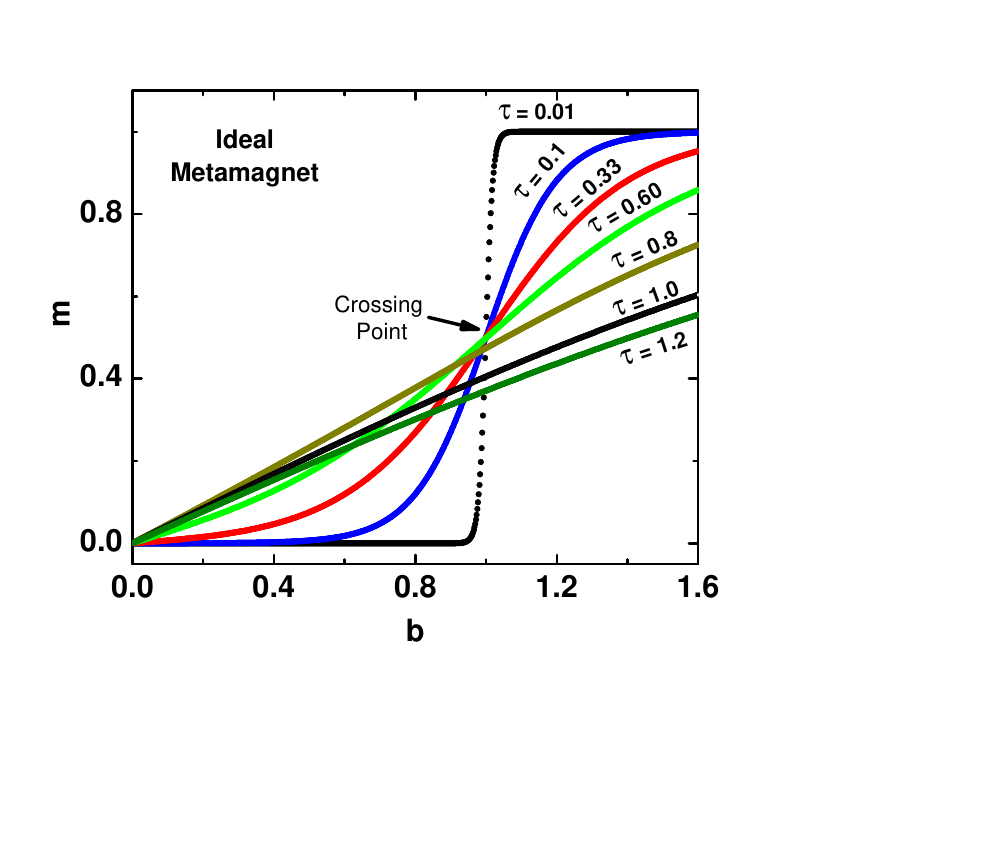}
\caption{\label{fig1} The magnetization isotherms of a minimal metamagnet. The magnetization at $T=0$ is zero unless the applied field is equal to or beyond the metamagnetic critical field $b=1$. There is thus a first order transition only at $T=0$.  At $T>0$ instead of a sharp jump there is a gradual rise in the magnetization albeit with a more rapid increase near the critical field resulting in a point of inflection.  Such a rapid rise completely vanishes (together with the inflection point) gradually as the temperature is raised.  $\tau$ is the reduced temperature, $\tau=kT/\Delta$, with $\Delta$ being the single energy scale in the model. }
\end{figure}

As seen in Fig.~2, the linear susceptibility is positive everywhere but the nonlinear susceptibilities change sign at characteristic temperatures. The third order susceptibility $\chi_3$ is negative for $\tau \geq 0.73$ while $\chi_5$ is negative for $\tau \geq 0.32$. We see that the peak temperatures $\tau_n$ for $\chi_n(\tau)$ are $\tau_1 = 0.68, \tau_3 = 0.27$ and $\tau_5 = 0.18$. The ratios of the peak temperatures are close to what is observed in experiments. Moreover, it seems that $\chi_3$ changes sign approximately where $\chi_1$ is maximum. Likewise, $\chi_5$ changes sign around where $\chi_3$ has a maximum.  Since the peaks of the $\chi_{n}$'s are of order 1, the measured susceptibilities $\gamma (\gamma/\Delta)^n \chi_{n}$ scale inversely with $\Delta^n$.

\begin{figure}
\includegraphics[width=110mm]{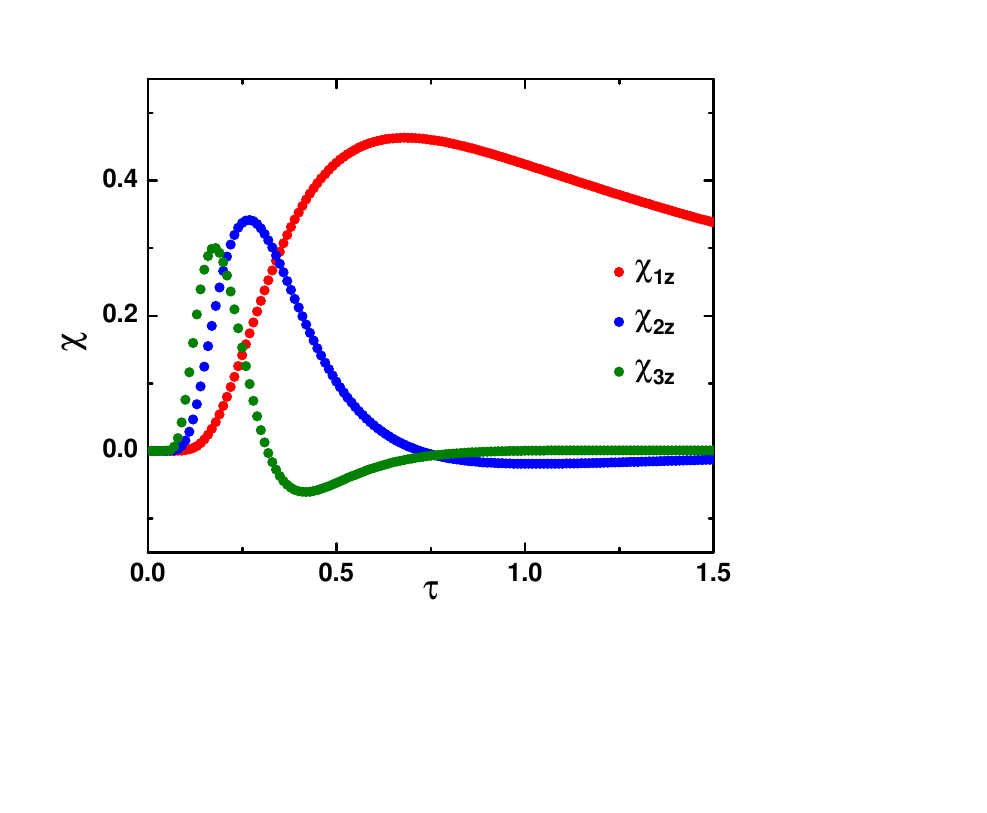}
\caption{\label{fig2} Shows the linear, $\chi_1$, third order, $\chi_3$, and fifth order, $\chi_5$, susceptibilities calculated in the minimal model as a function of the reduced temperature, $\tau =kT/ \Delta$, for the case when the magnetic field is parallel to the z-axis. Note that the positions in temperature of the maxima in the susceptibilities move to progressively lower values - reduced roughly by a factor of 2 going from $\chi_1$ to $\chi_3$ and from $\chi_3$ to $\chi_5$.}
\end{figure}

While the susceptibilities describe the low field behavior of the magnetization there are also notable features in high fields (Fig.~3). The peak in the magnetization seen at low fields shifts to a lower temperature and approaches the limit $T=0$ precisely for $b=1$. For constant fields $b > 1$ the behavior is Curie-like for most of the temperature range, but saturates at the lowest temperatures.\\

\begin{figure}
\includegraphics[width=110mm]{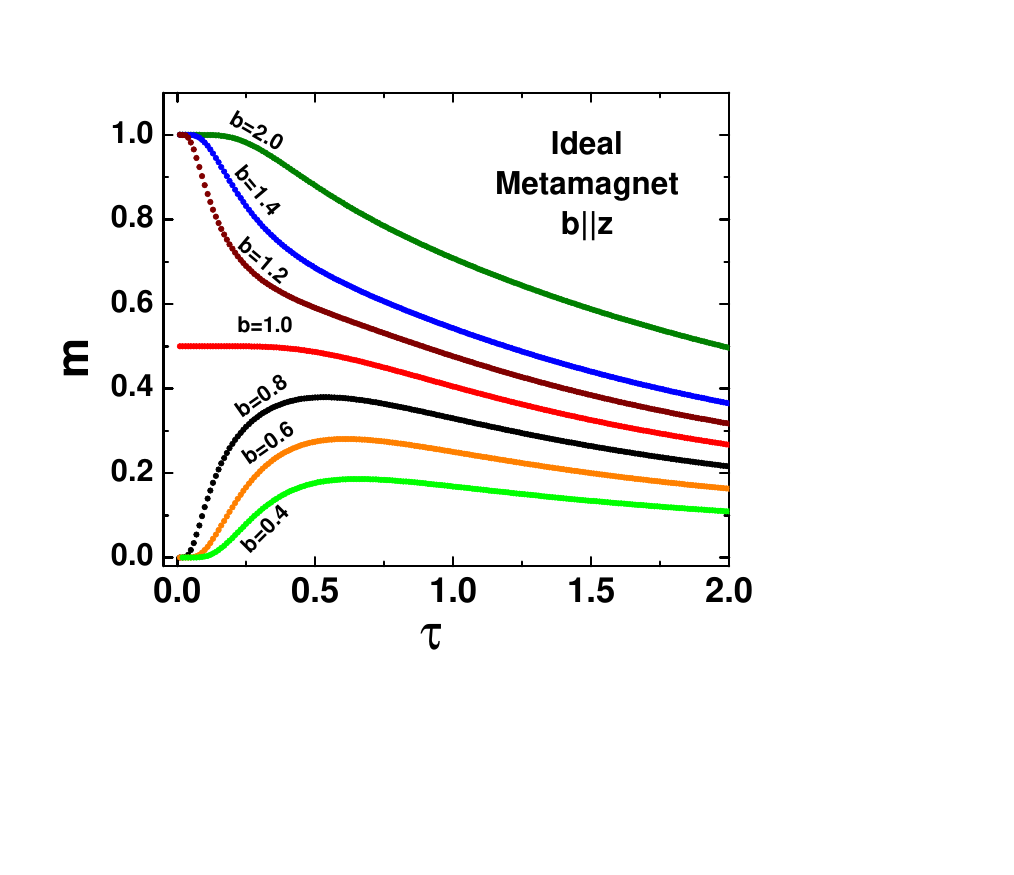}
\caption{\label{fig3} The temperature dependent magnetization of a minimal metamagnet at various constant magnetic fields both above and below the critical field. The magnetization at $T= 0$ is zero unless the applied field is equal to or beyond the metamagnetic critical field $b=1$. In measurements however a non-zero $m$ is observed \cite{PaulsenJLTP1990, AndreevPhilMag2003} for $b<1$.}
\end{figure}
\subsubsection{B perpendicular to z}
In this case, the energy levels are  $\epsilon_1 = \frac{1}{2}\Delta(1-r),  \epsilon_2 = \Delta$ and $\epsilon_3 = \frac{1}{2} \Delta(1+r)$, with $r=\sqrt{1+4b^2}$ (see Fig.~6b). 
The ground state is always $\epsilon_1$, which is paramagnetic at low $B$.

The partition function is 

\begin{equation} 
e^{-F/kT}=e^{-1/\tau } +2e^{-1/{2\tau }} \cosh {(r/2\tau) }   \label{eq:FforBx}
\end{equation}   

\noindent The magnetization is 
\begin{equation} 
m=\frac{4b}{r}\, \frac{\sinh {(r/2\tau)}}{e^{-1/2\tau}+2 \cosh{(r/2\tau)}}  \label{eq:mforBx}
\end{equation} 
and the susceptibilities in any direction perpendicular to $z$ are given by:
\begin{equation} 
\chi_{1x} = \frac{2a-1}{a+1} \label{eq:chi1forBx}
\end{equation} 
\begin{equation} 
{\chi }_{3x}= - \frac{4a\tau(2a+1)-10a-4\tau-1}{2\tau(a+1)^2}  \label{eq:chi3forBx}
\end{equation} 
\begin{equation} 
{\chi }_{5x}= \frac{3N_5}{4\tau^2 (a+1)^3} \label{eq:chi5forBx}
\end{equation} 
\noindent with $N_5 = 2a\tau(8a^2 \tau+12a\tau-10a-11)-6a^2+3a-8\tau^2-2\tau$. 
The magnetization evolves smoothly with field and has no discontinuities (Fig.~4). All of the susceptibilities, linear as well nonlinear, increase in magnitude monotonically with decreasing temperature, as shown in Fig.~5. 

\begin{figure}
\includegraphics[width=110mm]{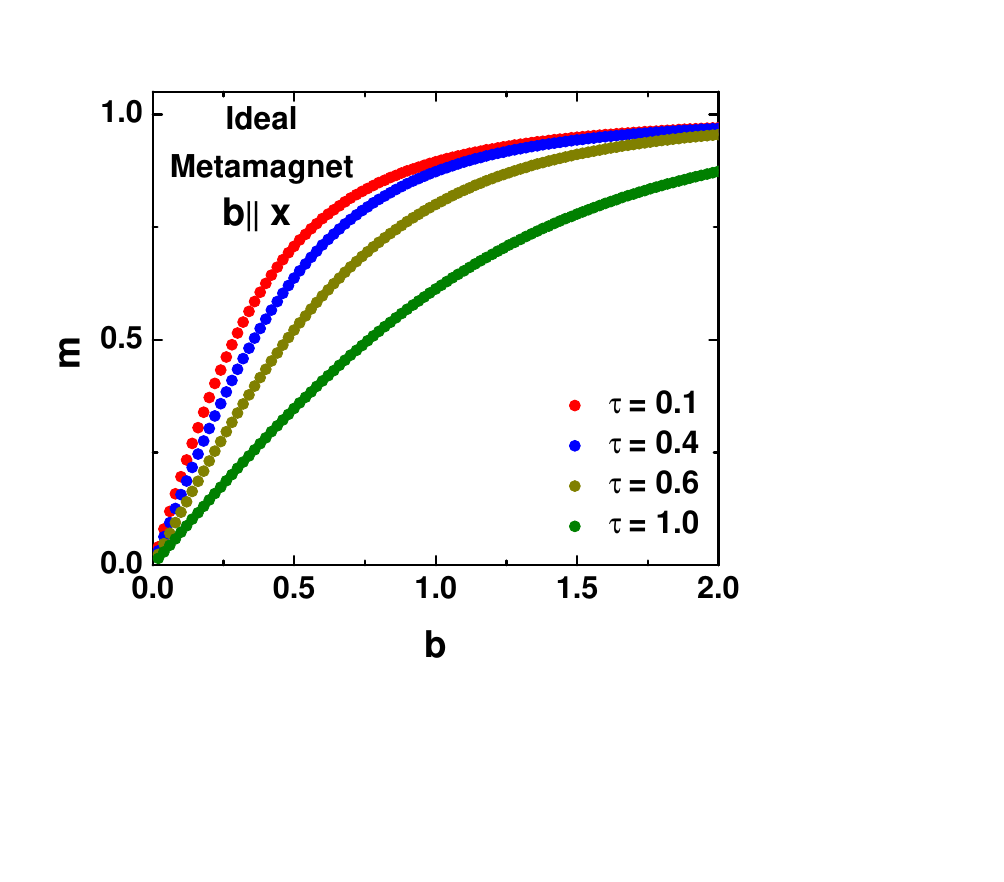}
\caption{\label{fig4} The magnetization isotherms for fields parallel to the x-axis i.e. perpendicular to the metamagnetic direction for four different temperatures. Although the behavior for this orientation appears similar to that of a paramagnet it is distinct as given by Eq.(9). }.
\end{figure}

\begin{figure}
\includegraphics[width=80mm]{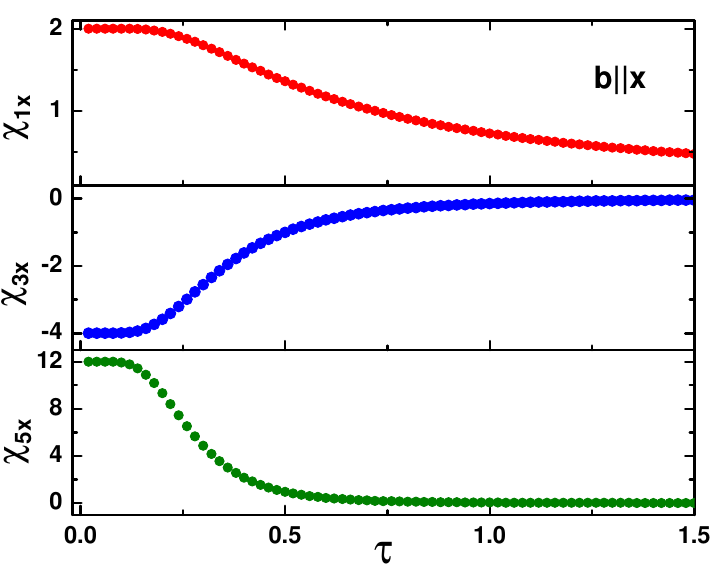}
\vspace{25mm}
\caption{\label{fig5}Shows the linear, $\chi_{1x}$, third order, $\chi_{3x}$, and fifth order, $\chi_{5x}$, susceptibilities calculated in the model as function of the reduced temperature, $\tau=kT/ \Delta$, for the case when the magnetic field is perpendicular to the metamagnetic direction. In this geometry the behavior of  $\chi_{1x}, \chi_{3x}$, and $\chi_{5x}$ is monotonic, with successive higher orders alternating from positive to negative, with no zero crossing for any of them. Note that $\chi_{1x}$ turns out to be significantly larger than $\chi_{1z}$ whereas the opposite is true in real systems.  This can be easily rectified by taking an anisotropic $\gamma$ or g-factor. There is direct experimental evidence for large g-factor anisotropies in many heavy fermion metamagnets \cite{SichelschmidtPRL2003, AltarawnehPRL2012}}.
\end{figure}
\subsubsection{B in the xz plane}
This is the general case, when the field is at an angle $\theta$ with the z axis. The Hamiltonian (\ref{eq:Hmin})   is conveniently written as
\begin{equation} 
\mathcal{H}=\ \Delta \left(S^2_z-  b_z\, S_z - b_x\,S_x \right) \label{eq:HforBxz}
\end{equation} 
\noindent The three eigenvalues of $\mathcal{H}$ are, $\epsilon_k =  \Delta \lambda_k$, where the $\lambda_k$ are the roots of the equation
\begin{equation} 
\lambda ^3-2 \lambda^2+{\lambda}\left(1-b^2\right)+b_x^2 =0 , \label{eq:EqforBxz}
\end{equation} 
 i.e., $\lambda_k =\frac{2}{3} \left(1+\sqrt{1+3b^2} \cos \dfrac{\alpha+2\pi k}{3}\right)$ for $k=1,2,3$. Here $\alpha$ is $\arctan (\sqrt{3}Y/X)$ for $X>0$ and   $\pi -\arctan (\sqrt{3}Y/|X|)$ for $X<0$, with 
$$X=6\,{b_z}^2 - 3\,{b_x}^2 -2/3$$ 
$$Y = \sqrt{4\, b_z^2 (b^2-1)^2+ b_x^2 (27\,b_z^2 +b^2+ 4\, b^4)}$$
These eigenvalues are plotted in Fig.~6 as a function of $b$ for several values of $\sin\theta = b_x/b$. It is seen that, for  $b$ near 1 and small  $\theta$, we have an avoided crossing described approximately by 
\begin{equation}
\lambda_{1,2} \approx \frac{1}{2}\left(1-b_z \pm \sqrt{(1-b_z)^2+2b_x^2}\right). \label{eq:AvoidedCrossing}
\end{equation} 

At any angle $\theta $, the magnetization in the direction of $\bm B$ is $\gamma m_\parallel =-\partial F/\partial B$. Fig.~7  has plots of $m_\parallel$ at $T=0$, showing that the effect of $\theta$ is similar to the effect of $T$ at $\theta =0$, shown in Fig.~1. For small $\theta$ and $\tau$, $m_\parallel$ rises sharply near $b = 1$, and $\partial m_\parallel / \partial b$ has a peak of height 

\begin{equation}
p = \frac{\sqrt {2}}{4b_x}\tanh \left( {\frac {\sqrt {2}b_{{x}}}{\tau}} \right)
\end{equation}
and width $1/p$. This reduces to $2 \tau$ for $b_x = 0$, as already noted after Eq.~(\ref{eq:mforBz}),  and to $2\sqrt{2}\,b_x$ for $\tau=0$.
 
\begin{figure}
\includegraphics[width=80mm]{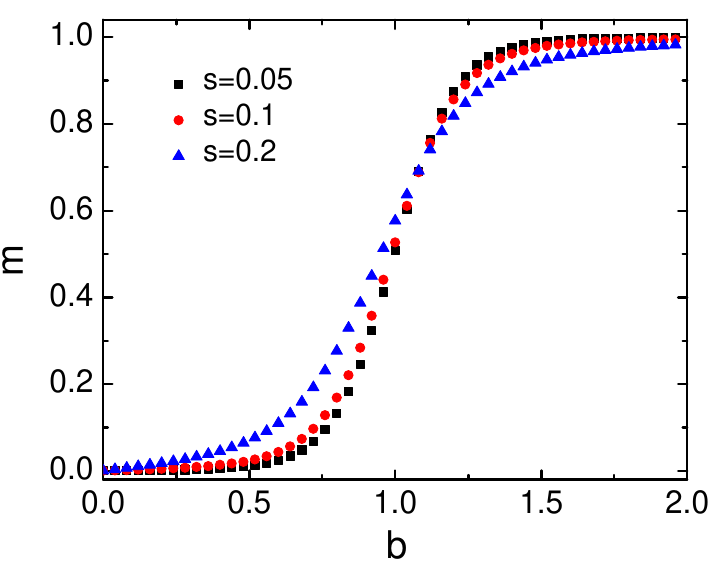}
\vspace{20mm}
\caption{\label{fig6}Shows the magnetization calculated in the model as a function of the reduced magnetic field, $b$, applied in a
 direction tilted by angle $\theta$ away from z, with s=sin$\theta$. }
\end{figure}

We can also obtain the low-field susceptibilities for any direction of the field, using the expansions: 
$$\lambda_1 = b_x^2 \left(-1 + b_x^2 - 2 b_x^4 +(4b_x^2-1)b_z^2- b_z^4 \right)$$
$$\lambda_{2} = 1- \frac {\lambda_{1}}{2} - \frac{Y}{2} \left(21\,b_x^{4} -12\,b^{2}b_x^2+b^4-4b_x^{2}+b^2+1 \right)$$
and $\lambda_{3}$ equal to $\lambda_{2}$ with $Y \rightarrow -Y$.

\begin{figure}
\includegraphics[width=100mm]{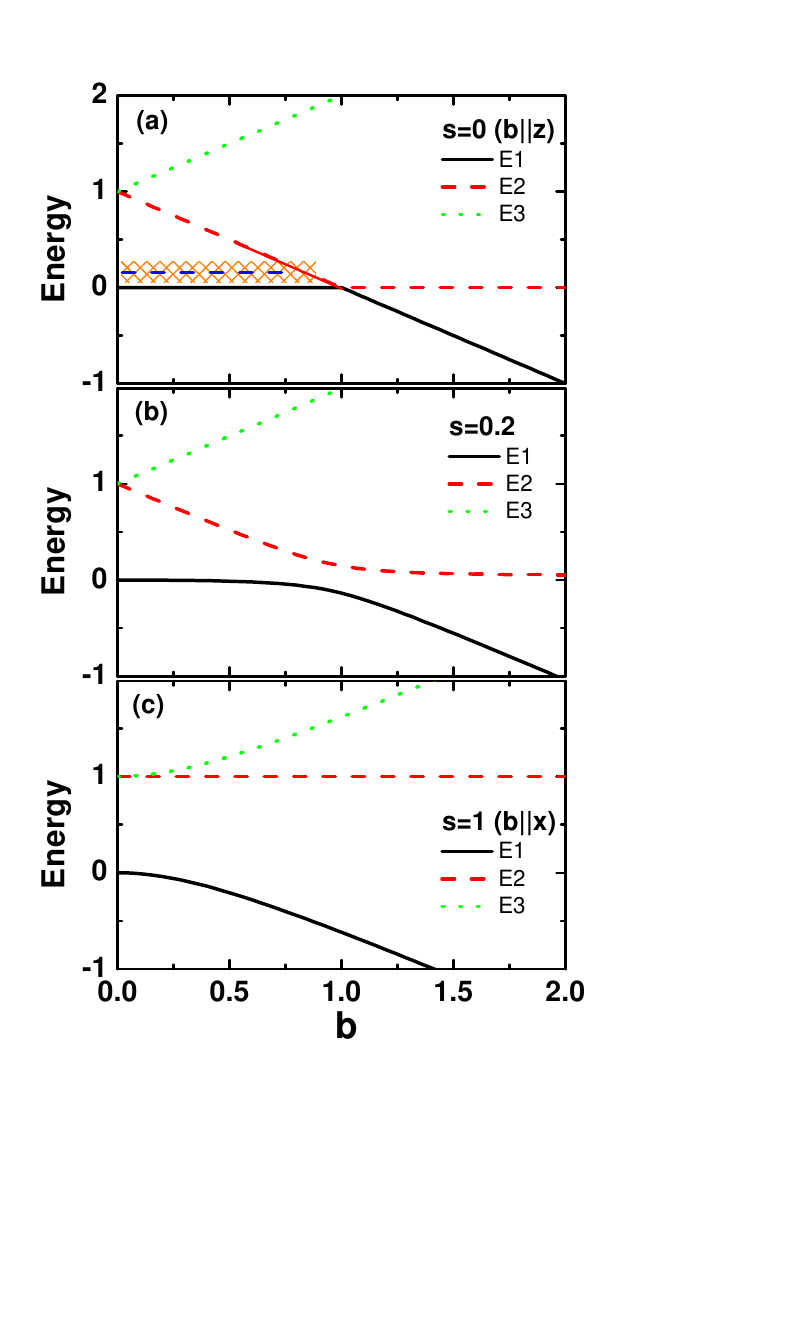}
\vspace{-15mm}
\caption{\label{fig8}Shows the energy levels calculated in the model as a function of the reduced magnetic field, $b$, for (a) when the magnetic field is tilted by an angle $\theta = sin^{-1}s$, (b) parallel to the metamagnetic direction $z$ and (c) perpendicular to the metamagnetic direction. Also shown in (b) is the scenario when the ground state has a non-zero width (hashed region) and is a narrowly split doublet (horizontal dashed line).}
\end{figure}

To first order, $m_z/b_z=\chi_{1z}$ and $m_x/b_x=\chi_{1x}$ are given by Eqs.~(\ref{eq:chi1forBz}) and (\ref{eq:chi1forBx}), and $\chi_{1\parallel} = m_\parallel/b = \chi_{1z} c^2 + \chi_{1x} s^2$, with $c=\cos{\theta}$, $s=\sin{\theta}$. 

The third-order terms in $m_z/b_z$ are $\chi _{3z}b_{z}^{2}+\chi_{zx}b_{x}^{2}$, with $\chi_{3z}$  given by Eq.~(\ref{eq:chi3forBz}) and  $$\chi_{zx}=\frac{4\,{a}^{2}{\tau}^{2}+2\,a{\tau}^{2}-2\,{\tau}^{2}-2\,\tau\,a-2\,\tau-3}{(a+1)^{2}}.$$ 
\noindent
The result for $m_x$ is analogous, with $\chi_{3x}$  given by Eq.~(\ref{eq:chi3forBx}) and  $\chi _{xz}=\chi _{zx}$.  Then $\chi_{3\parallel}= \chi_{3z}c^4+\chi_{3x}s^4+2\chi_{zx}c^2 s^2.$

The fifth order terms are $\chi _{5z}b_{z}^{4}+\chi_{zzx} b_{z}^{2} b_{x}^{2}+\chi_{zxx}  b_{x}^{4}$ and the same for $z\rightleftarrows x$, with $\chi _{5z}$ and $\chi _{5x}$ from Eqs.~(\ref{eq:chi5forBz}) and (\ref{eq:chi5forBx}). Since $\chi _{zzx}=2 \chi _{xzz}$ and $\chi _{xxz}=2 \chi _{zxx}$, we still need only
\begin{eqnarray*}
\chi _{zxx} &=&\frac{1+6\tau +24\tau ^{2}-6\tau ^{2}(4\tau -1)(2a-1)}{12\tau
^{4}(a+1)}-\frac{Q_{zxx}}{8\tau } \\
\chi _{xzz} &=&\frac{48a\tau ^{4}-24\tau ^{4}-24\tau ^{3}-12\tau ^{2}-4\tau
-1}{12\tau ^{4}(a+1)}-\frac{Q_{xzz}}{8\tau }
\end{eqnarray*}
where $Q_{zxx}=$ $2\chi _{1z}\chi _{xx}+\chi _{1z}\chi _{1x}^{2}+4\chi
_{zx}^{{}}\chi _{1x}$ and   $z\rightleftarrows x$ gives $Q_{xzz}$.

Then $\chi_{5 \parallel}=\chi _{5z}c^{6}+3\chi _{xzz}c^{4}s^{2}+3\chi_{zxx}c^{2}s^{4}+\chi _{5x}s^{6}.$\\

  
\subsection{Specific Heat} 

The specific heat at constant volume is given by $C_V(B,V,T)/T= - \partial^2 F/\partial T^2$, per formula unit, in addition to the non-magnetic contribution. The experiments are carried out at constant $P$, and $C_P$ differs from $C_V$ in the presence of magnetostriction, as we discuss in Subsection D. \\ 
For a B-field in the z direction, $C_V/T$ has a rich structure. At low fields the minimal model, from Eq.~(\ref{eq:FforBz}), predicts a peak at a temperature that scales with $\Delta$ in a manner similar to that seen in many HFS \cite {BrodalePRL1986, VollmerPRL2003}. At low $T$, the predicted field dependence  has the “M” shape seen in the inset of Fig.~9. A similar shape of  $C/T$ is indeed observed in some metals \cite{AokiJMMM1998}, and a nearly perfect M is seen in molecular magnets, where similar Hamiltonians are employed to describe the observed properties \cite{EvangelistiJMatChem2006}.  However, many HFS exhibit a single peak \cite{MullerPRB1989} at the critical field.   The extinction of $C_V/T$ at $b=1$ originates from the perfect crossing of the energy levels. This extinction can be lifted by an “anticrossing” which can be modeled through various means. The main part of Fig.~9 shows that this can be achieved with a small tilt of the magnetic field away from z.


\begin{figure}
\includegraphics[width=110mm]{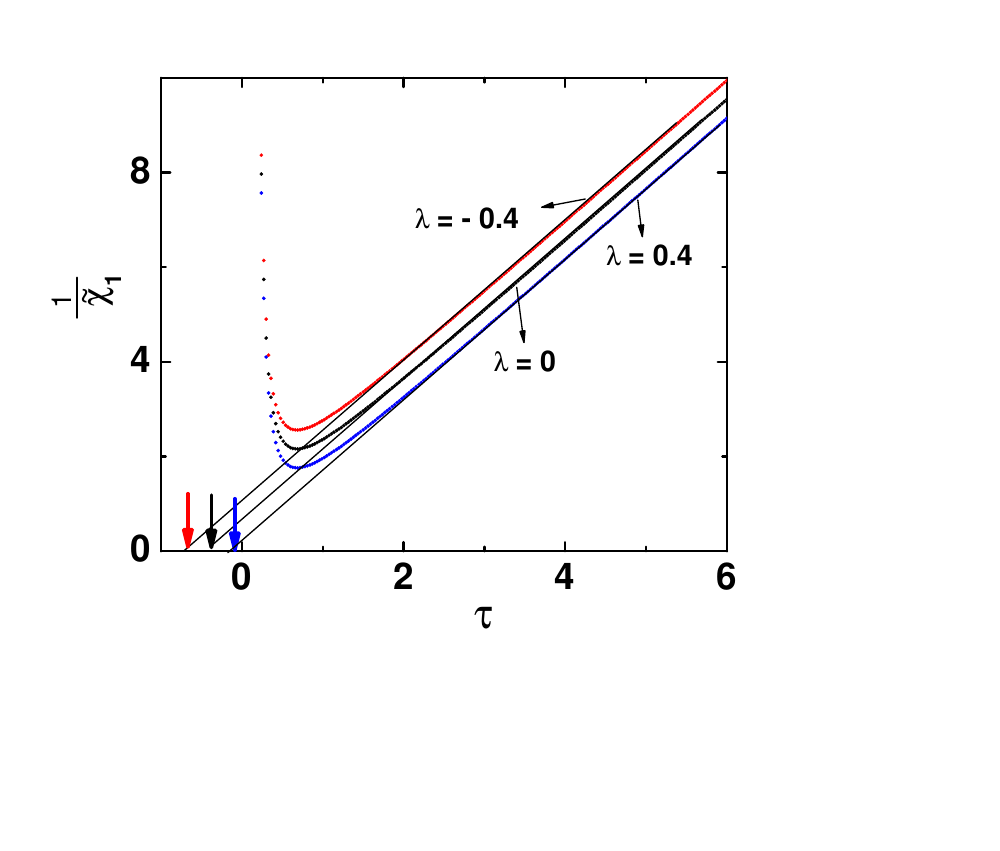}
\caption{\label{fig8}Shows the inverse susceptibility calculated in the model as a function of temperature. The three lines correspond to the three different values of the mean field parameter, $\lambda$ used in Fig.~8. A ferromagnetic mean field not only shifts the
 metamagnetic transition to lower fields it also sharpens it. The opposite is true for an antiferromagnetic mean field.}
\end{figure}

\subsection{Magnetoelastic Properties}
The magnetic free energy $F(B,T,V)$ depends on $V$ (and more generally on strain) through the energy scale $\Delta$,  and also through $\gamma$, which is proportonal to the g-factor. We discuss in detail the effect of  $\Delta(V)$;
the effect of $\gamma(V)$ is treated in the same way, noting also that $\partial F /\partial \gamma = - Bm$.\\
To a sufficient approximation, the pressure is
\begin{equation} 
P = -K_0\, \frac{V-V_0}{V_0} - \frac{\partial F }{\partial V}, 
\label{eq:Pressure} 
\end{equation}  
where $K_0$ and $V_0$ are effectively constant and $\partial F /\partial V = \beta \eta$, with
\begin{equation} 
\beta = \frac{ d \Delta} {dV}, \quad \eta =\frac{\partial F }{\partial \Delta}.
\label{eq:beta}
\end{equation}

\noindent 
We note that $\beta(V)$ is a property of each material, while $\eta(\Delta,\bm B,T)$ is universal, within the model.\\
The bulk modulus is $ K=- V\, \partial P /\partial V$ with
\begin{equation} 
\frac{\partial P}{\partial V} = - \frac{K_0}{V_0} - \frac{\partial^2 F }{\partial V^2}.
\label{eq:K}
\end{equation}

The square of the longitudinal sound velocity is proportional to $K$, hence its dependence on B and T is
\begin{equation} 
\frac{\delta {v_s}}{v_s}=\frac{1}{2}\, \frac{\delta {K}}{K} \propto  \frac{\partial^2 F }{\partial V^2} =\eta \frac{d\beta}{dV}+\beta^2\, \frac{\partial\eta}{\partial \Delta}
\label{eq:vs}
\end{equation}  

The field dependence of the sound velocity for $\bm B$ parallel and perpendicular to the metamagnetic direction is illustrated in Fig.~10, for $\Delta d\beta/dV=0.4 \, \beta^2$ . In the top panel, there is an asymmetric dip at $b=1$ that sharpens as $T\rightarrow 0$.   The dip comes from $\Delta\, \partial\eta/\partial \Delta$ and the asymmetry comes from $\eta$, with 
\begin{equation} 
\eta = \frac{1}{\frac{1}{2}e^{\mathrm{\Delta }\mathrm{/kT}}{\mathrm{sech} \left(\gamma B/kT\right)+1\ }}
\label{eq:eta}
\end{equation} 
and  $\partial\eta/\partial \Delta=(1/kT)(\eta(1-\eta)$.
  The behavior shown bears a striking resemblance to the experimentally observed sound velocities \cite {LuthiJMMM1987, WolfJLTP1994, YanasigawaJPSJ2002, FellerPRB2000, SuslovJLTP2000}.  \\
The temperature dependence of the sound velocity is shown in Fig.~11 and exhibits the characteristic dip at a temperature of the order of $\Delta/k$ observed experimentally.\\

Eq.~(\ref{eq:Pressure}) also provides access to magnetostriction. We have $dP =-(K/V) dV - \beta\, \partial \eta/\partial B$, hence
\begin{equation}
 \frac {1}{V}\,\left.{\frac{\partial V}{\partial B}}\right|_P = \frac{\beta}{K}\frac{\partial \eta}{\partial B}= \frac{1}{K}\frac{\partial M}{\partial V}, \label{eq:magnetostriction}
\end{equation}
with $ \partial M/\partial V = \beta \,\partial M/\partial \Delta  $.
 The measured values of the magnetostriction\cite {PuechJLTP1988} along z bear a striking resemblance to this result with $\eta$ from Eq.~(\ref{eq:eta}) or $M$ from Eq.~(\ref{eq:mforBz})  .\\


\begin{figure}
\includegraphics[width=110mm]{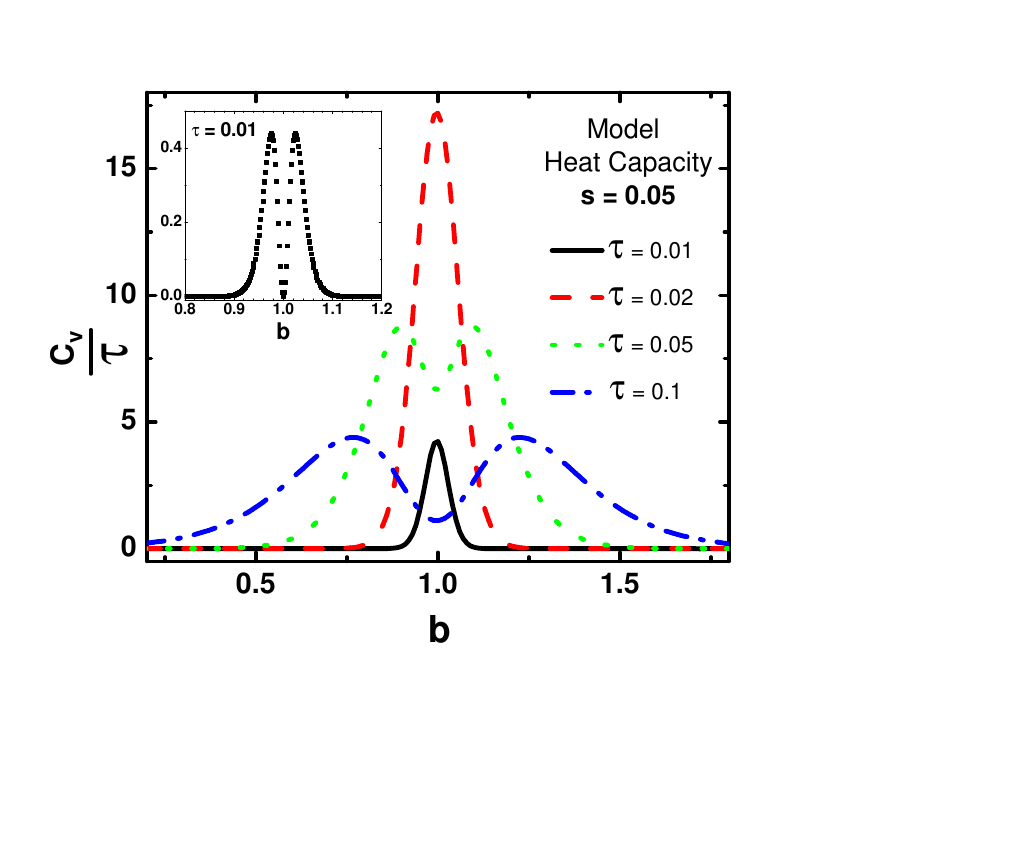}
\caption{\label{fig9} Shows the heat capacity as a function of the magnetic field at different temperatures calculated for a small tilt ($s$\,=\,0.05) of the magnetic field with respect to the z-axis.  Note the single peak in $C_V/T$ at the lowest temperature.  Without this tilt, or in general when there is no anti-crossing of the energy levels, the field dependence of the heat capacity has a double peaked structure at very low $T$, as shown in the inset.  Behavior qualitatively similar to that shown in this figure has been observed in CeRu$_2$Si$_2$ \cite{Japanese}}.
\end{figure}

\begin{figure}
\includegraphics[width=100mm]{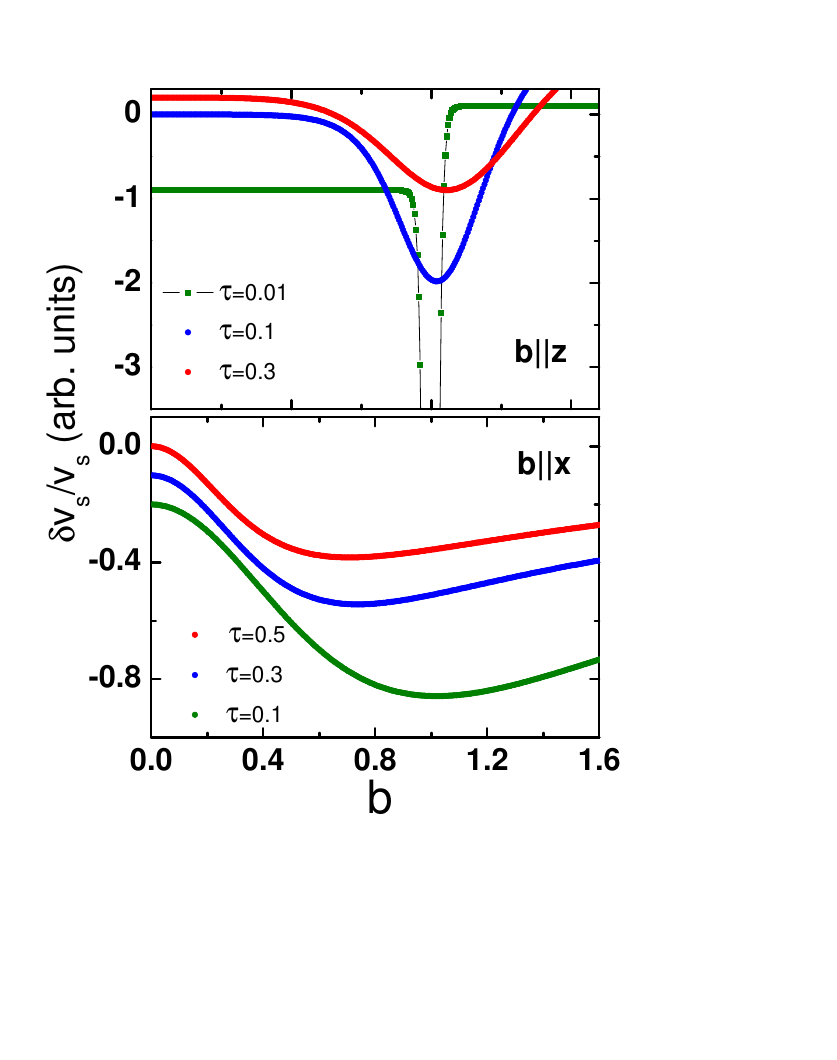}
\caption{\label{fig10} Shows the expected behavior of the longitudinal sound velocity as a function of the magnetic field at different temperatures obtained in the model.  The top panel illustrates the behavior expected when the field is parallel to z and the bottom panel when it is perpendicular to it.}.
\end{figure}

\begin{figure}
\includegraphics[width=100mm]{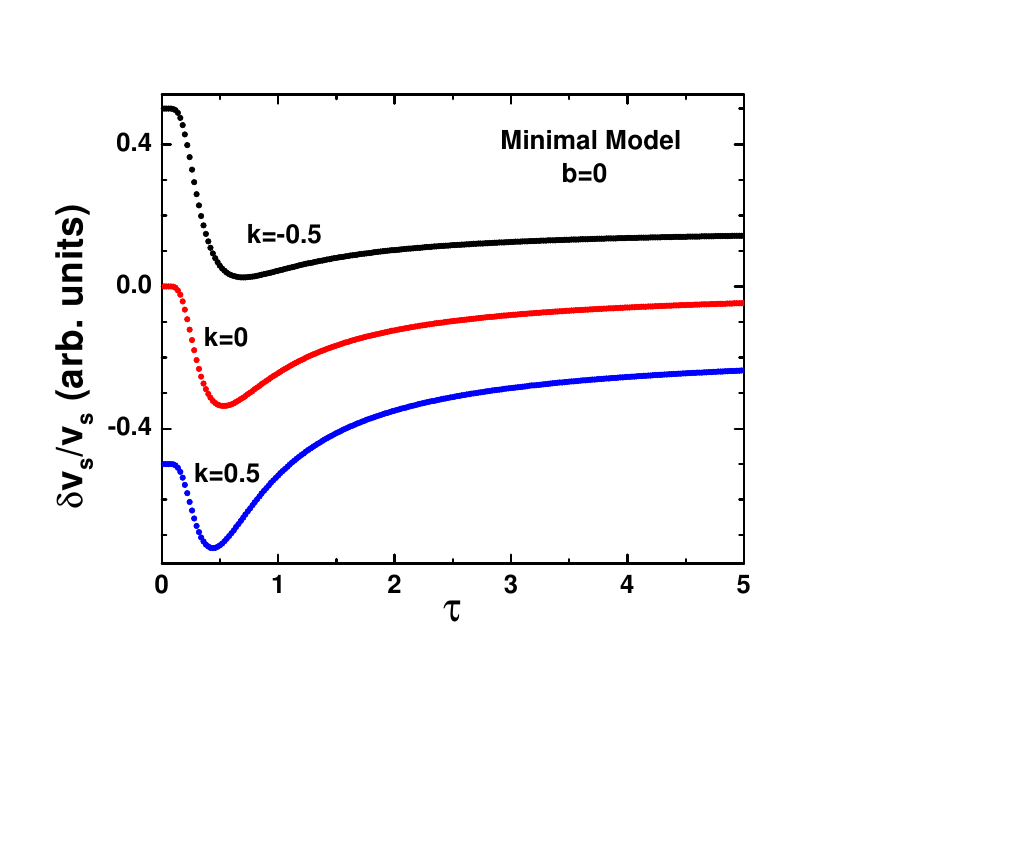}
\caption{\label{fig11} Shows the temperature dependence of the longitudinal sound velocity at zero field obtained in the model.}
\end{figure}


Apart from the above other useful relations also follow. There is a general relationship between derivatives of the free energy $F$ with respect to $\Delta$ and those with respect to $B$. For example:

\begin{equation} 
\gamma^2 \frac{\partial F\ }{\partial \Delta }={\left(\frac{\partial F}{\partial B}\right)}^2-kT\frac{{\partial }^2F}{\partial B^2}\ ,
\end{equation} 

which also becomes, with $P_0 = (K_0/V_0)(V-V_0)$, 

\begin{equation} 
P-\ P_0=\ -\left(\frac{\partial \Delta }{\partial V}\right)\,\left[\frac{kT}{\Delta}\chi \left(T,B\right)+\ {m(T,B)}^2\right] .
\end{equation}

Here $m$ and the susceptibility are both fully dependent on the magnetic field. \\

\subsection{Observables}
While the experiments measure samples at constant pressure, all theory is indeed done at constant volume. In the absence of magnetostriction, the constant pressure and constant volume observables are identical. Otherwise they are related by  thermodynamics.  To discuss them, it is convenient to revert to the standard unscaled variables.\\
\indent The dynamic susceptibility is $\chi(B,V)=\partial M/\partial B$ = $(\gamma^2/\Delta) \, \partial m/\partial b$. We have
\begin{equation} 
{\chi }_P-{\chi }_V= \left( \frac{\partial M}{\partial V} \right) \left. {\frac{\partial V}{\partial B}}\right|_P = \frac{V}{K}\left( {\frac{\partial M}{\partial V}} \right)^2.
\label{eq:chiP}
\end{equation} 
As in Eq.~(\ref{eq:magnetostriction}), $ {\partial M/\partial V} =
\beta\, {\partial M/\partial \Delta}$ and ${\partial M / \partial \Delta}$ is strongly peaked at the critical point,  differing from $(1/\gamma)\, {\partial M / \partial B}$ by terms of order $e^{-\Delta/kT}$.\\
For the specific heat, analogously to Eq.(\ref{eq:chiP}),
\begin{equation} 
C_P-C_V= T \left( \frac{\partial S}{\partial V} \right) \left. {\frac{\partial V}{\partial T}}\right|_P = \frac{TV}{K}\left( {\frac{\partial S}{\partial V}} \right)^2,
\label{eq:CP}
\end{equation}
with $\partial S/\partial V =  \partial P/\partial T = \beta \, \partial \eta/\partial T $.
Similar to the susceptibility the correction strongly peaks at the critical point. \\
Similarly for the bulk modulus,


\begin{equation} 
 \frac{K_M}{K_B}\ =\frac{{\kappa }_B}{{\kappa }_M}\ =\frac{{\chi }_P}{{\chi }_V }>1 
\end{equation} 

\noindent Here $\kappa_H$ is the compressibility at constant magnetic field, and $\kappa_M$ is that at constant magnetization. We note that in the classical analysis of the temperature dependence of the sound velocity in air Newton had used the isothermal compressibility. But Lagrange noted that at a typical sound frequency, it should be the adiabatic compressibility that should be used in the expression for sound velocity. In solids the difference between the adiabatic and isothermal compressibilities is usually unimportant.  However, here we have the additional question :  should it be the constant “magnetization” or constant field compressibility that should be used to calculate the sound velocity.\\

 While the discussion above considered the difference between the susceptibilties at constant volume and constant pressure, we must also evaluate the effect of such a transformation on the magnetization per se.  The magnetization at constant volume is of course given by $- (\partial F/\partial B)_V$.  But the experimental volume $V$ is a function of both $P$ and $B$, and in general also $T$.  Confining ourselves to an isothermal situation $V=V(P,B)$. Since the volume depends on the magnetic field for each field we can seek that pressure which would restore the volume back to its zero field value.  Alternately, since the single energy scale is pressure dependent we can seek that value of $ \Delta$ which would restore the volume.  In effect, to compare theory with experiment  
$M$ has to be evaluated  at $\Delta_{0}-K/V\,\beta^{2}(\partial F/\partial\Delta))$.  An example plot of such a conversion from constant V to constant P is given in Fig.~12 and may be compared to a similar analysis given by Matsuhira et al \cite{MatsuhiraJPSJ1997}.

\begin{figure}
\includegraphics[width=100mm]{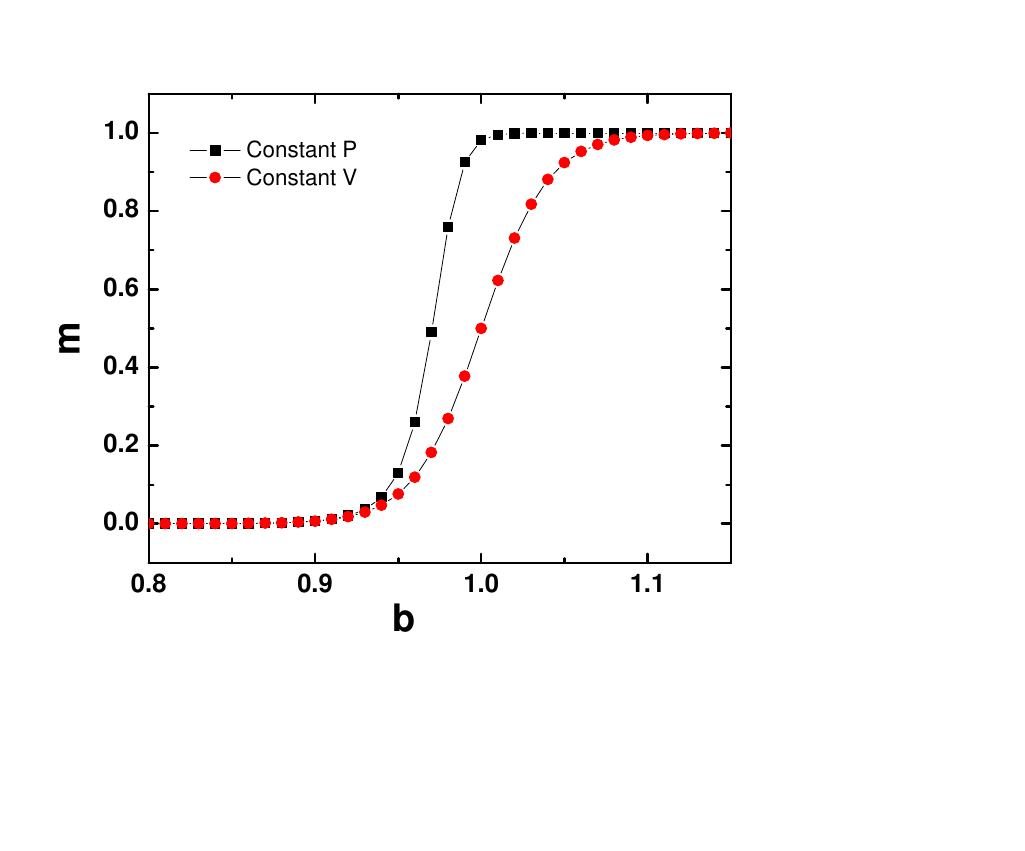}
\caption{\label{fig12} Shows the transformation of the magnetization from a constant $P$ to a constant $V$ condition.  For the purpose of illustration here $ (V/\Delta)\, d\Delta/dV = 0.4$.}
\end{figure}

\subsection {Ginzburg Landau Description} 
A magnet is often described by a Ginzburg-Landau free energy
which in our dimensionless notation is \\
\begin{equation}
F_{GL}= -b\,m+ a_2 m^2 + a_4 m^4 + a_6 m^6+...
\end{equation},   \\
Here the coefficients $a_i$ (i = 2, 4, 6) incorporate the temperature dependence. The equilibrium equation of state $b(m,T)$ then is given by the minimum of $F_{GL}$ with respect to $m$. On the other hand, at small $b$ \\

$m= \chi_1 b + \chi_3 b^3 + \chi_5 b^5+...$. \\

We see that \\

$a_2 = 1/2 \chi_1 $ \\

$a_4 = {-\chi_3 }/{4 \chi_1^4 }$  \\

$ a_6= - \left(\chi_5/[6 \chi_1^6 \right) \left[1-3 \chi_3^2/{\chi_5 \chi_1 }\right]$  \\

These are model independent relationships. All coefficients of $F$ are positive if $\chi_1$ is positive and $\chi_3$ and $\chi_5$ are both negative. This is the case at high temperatures. The curious result here is that whenever $\chi_5 > 0$, then (for $\chi_1 >0$) the parenthesis must be negative, or $3 (\chi_3)^2 > \chi_5 \chi_1$ for the equilibrium state to be bounded. 
In the interesting range of temperatures in UPt$_3$ where $\chi_3 >0$ and $a_4 < 0$, the coefficient $a_6 < 0$, indicating an overall instability of the GL expansion. In any case GL is supposed to be an expansion for small order parameters and metamagnetism leads to a large magnetization. Therefore one might not expect GL to be a valid representation.\\

\section{INCLUDING OTHER DEGREES OF FREEDOM}

\subsection{Mean Field Effects} 
The models above are all single site models. In a solid there are interionic exchange effects that will affect metamagnetism.  These effects can be incorporated in the mean field approximation as noted by Morin and Schmitt \cite{MorinSchmittPRB1981}. In this section we show how they modify two salient features of metamagnetism: the magnetization isotherms near the critical point (see Fig.~1) and the temperature dependence of the susceptibilities (Fig.~2).  To be definite, the discussion will be within the framework of the minimal model with B-field along z (Section IA), although some of the results are easily extended to other scenarios.   In the mean field approximation, the exchange interaction $ \sum J_{ij} \bm{S}_i \cdot \bm{S}_j$ (for $i \neq j$) is accounted for by the prescription: the external field $B$ everywhere is replaced by $B + \lambda M$, and $M$ is computed self-consistently.  Many well-known spin arrangements can result, but we will confine our treatment to the simplest cases: ferromagnetism and antiferromagnetism. If the generalized $J_{ij}$ depends on the relative position $\bm{u} = \bm{r}_i - \bm{r}_j$, it follows that in terms of its Fourier transform $J(\bm{q})$, $\lambda = J(q=0)$ leads to ferromagnetism if it is positve,  and for nearest neighbor interaction, $\lambda \approx zJ$ with $z$ as the coordination number. For antiferromagnetism, we must have a negative $\lambda = J(q = G/2)$, where $\bm{G}$ is the appropriate reciprocal lattice vector. We deal only with cases where the mean field alters the metamaghnetic state but does not lead to an ordered phase.\\
\subsubsection {Magnetization - Shift of $B_c$}
Recall the characteristic metamagnetic behavior, shown in Fig.~1: when $T\rightarrow 0$ the magnetization rises abruptly at the critical field $b=1$, and the isotherms cross at $m= 1/2$. This behavior persists when $\lambda$ is finite, but not too large in magnitude, with one main difference: For ferromagnetic exchange ($\lambda>0$),  the critical field shifts from $b=1$ to a smaller value and the metamagnetic transition is sharpened; for antiferromagnetic exchange the contrary happens. The shifted critical field is at $B_c = \Delta (1-\lambda/2)$ at $T=0$ and remains close to it for small $\tau$. 
All these features are seen in Fig.~8, where the arrows show the position of $b_c=B_c / \Delta$. \\
 More in detail, the self consistency equation is $\tilde{m}=m(\tilde{b})$, where $\tilde{b} = b + {\lambda}\tilde{m}$ and the function $m(b)$ is given in Eq.~(\ref{eq:mforBz}). The critical point is at $(\tilde{b}=1, \tilde{m}=1/2)$, which gives $b_c = b+\lambda / 2$. The width of the transition, $\tilde{w}$, is given by the inverse of $\partial \tilde{m}/\partial b$ at the critical point, where it peaks.  Using $\partial \tilde{m}/\partial b =(\partial {m}/\partial b )(1+\lambda \, \partial \tilde{m}/\partial b)$ we see that in general $\tilde{w}$ scales as $1-\lambda$; for small $\tau$, $\tilde{w}=\tau(1-\lambda)$.

\subsubsection {Susceptibilities}
In the mean field approximation, the linear and nonlinear susceptibilities become \\
 
\begin{equation} 
{\widetilde{\chi }}_1=\frac{{\chi }_1}{1-\lambda {\chi }_1} \label{eq:chi1tilde}
\end{equation} 

\begin{equation} 
{\widetilde{\chi }}_3=\ \frac{{\chi }_3}{{(1-\lambda \ {\chi }_1\ )}^4}  \label{eq:chi3tilde}
\end{equation}

\begin{equation} 
{\widetilde{\chi }}_5=\ \frac{{(\chi }_1{\chi }_5\ -3{{\chi }_3}^2)\lambda +{\chi }_5}{{(1-\lambda \ {\chi }_1\ )}^7}, \label{eq:chi5tilde}
\end{equation} 

\noindent where $\chi_1$, $\chi_3$, $\chi_5$ are the susceptibilities for $\lambda=0$, along $z$ or along $x$ . These are general formulae, valid as long as $| \lambda | \chi_1$ is less than 1, i.e.,  for temperatures above a (possible) ordering phase transition.\\

It is worth noting that $1/\chi_1$ and ${1/\widetilde{\chi }}_1$ share a characteristic temperture dependence, shown in Fig.~9.  They both have a minimum at the same temperature, $\tau_1 = 0.6835$. For $\tau > \tau_1$ we have 
 $\chi_1 \approx 2/(3\tau+1)=2\Delta/(3kT+\Delta)$. This looks like the susceptibility of an antiferromagnet with a Curie-Weiss temperature $-\Delta/3k$, i.e. a N\'eel temperature $\Delta/3k$. This behavior should not be interpreted as due to interionic exchange. However, it modifies the Curie-Weiss temperature for ${\widetilde{\chi }}_1$. From 
$1/{\widetilde{\chi }_1}=1/{\chi _1} - \lambda  \approx  (3\tau+1-2\lambda)/2$, which gives an effective Curie-Weiss temperature $ \Theta = (\Delta/3k)(2\lambda-1) = (T_1/2.05)(2\lambda-1)$.

\begin{figure}
\includegraphics[width=90mm]{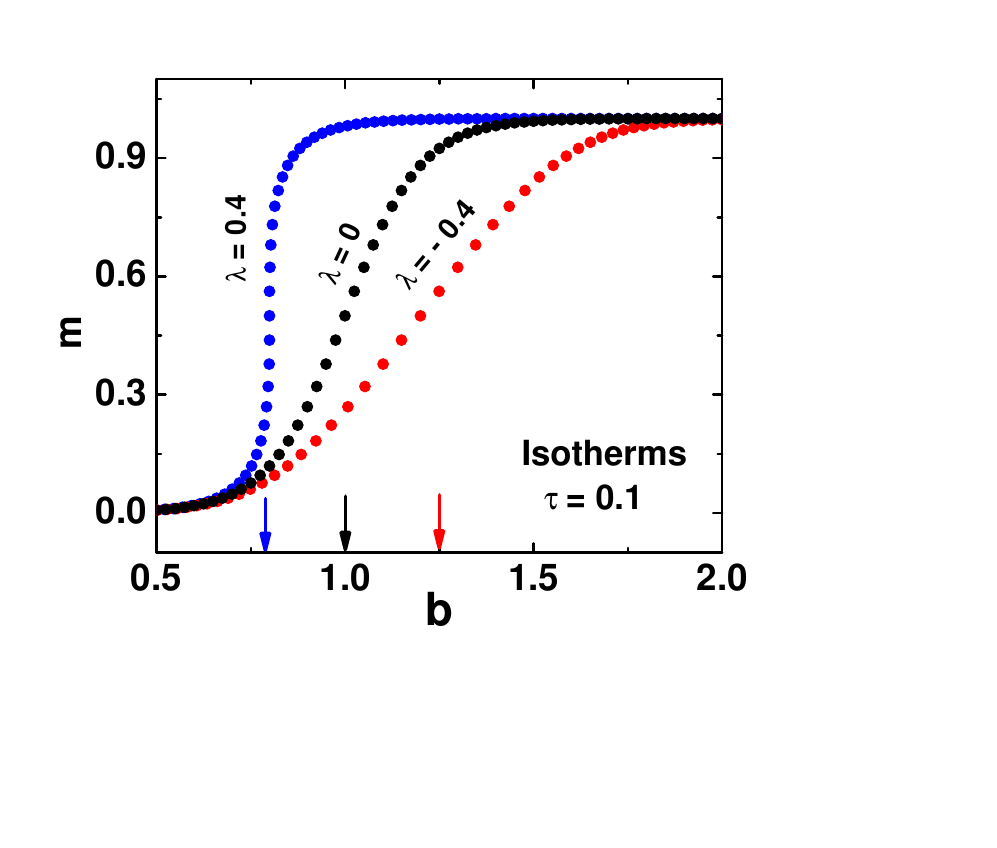}
\caption{\label{fig13}:  The magnetization isotherms with the inclusion of both ferromagnetic ($\lambda>0$) and antiferromagnetic ($\lambda < 0$) mean fields compared with the case of zero mean field.}.
\end{figure}

While the position of the peak in the susceptibility is not altered with the introduction of a mean field, the critical field is indeed sensitive to $\lambda$. As we have discussed above, positive values of $\lambda$ i.e. a ferromagnetic mean field shift $B_c$ to lower values and vice versa. Thus in those metamagnets which are close to a ferromagnetic instability the critical field can be sufficiently reduced to require a modified single energy scaling. Indeed in such materials as UCoAl the critical field can be a tenth of the expected field, $B_c=kT_1/\gamma$, based on linear scaling with $T_1$ \cite{SechovskyPhysica1986, AndreevSovPhys1985}. Figure 13 illustrates this point.  \\

\subsection{Level Broadening and Interactions}
While it is clear from the above discussion that the minimal model captures a number of salient features of itinerant metamagnets there are others that it is not able to account for.  Notable amongst these is the failure to account for the large non-zero values of the linear susceptibility at $T=0$ seen in almost all metamagnets.  While this is an obvious shortcoming relevant at zero-field there are also observations in the critical region that the model does not account for.  For instance, the differential susceptibility at $b=1$ follows a strict $T^{-1}$ relationship in the model for $T \rightarrow 0$. Experimentally as noted in a number of cases the differential susceptibility saturates in the subKelvin region \cite{FlouquetPhysicaB1995} as does the longitudinal sound velocity \cite{FellerPRB2000}.  To account for this behavior we augmented the minimal model with a energy level broadening, shown by a hatched region in Fig.~7b, parametrized by $w$ and by replacing the temperature $T$ by $ \sqrt{T^2+w^2}$. Numerically the value of $w$ turns out to be the temperature where a deviation from a linear behavior in the $ (\partial m/ \partial b)^{-1}$ vs. T plot is seen.  Such a plot appropriate for CeRu$_2$Si$_2$ is shown in Fig.~14 where $w=0.6$ K. The same value of level broadening is also able to account successfully for the critical field behavior of the magnetization (see Fig.~15) as $T \rightarrow 0$ and the zero field longitudinal sound velocity in the same temperature limit \cite{ShivaramPRB2015}.  Experimentally one observes a near $T^2$ dependence for both the magnetization \cite{PaulsenJLTP1990} (for $B$ close to $B_c$) and the zero field sound velocity at mK temperatures \cite{BatloggPRB1986}.  The minimal model however yields a flat or T-independent behavior for both the physical quantities.

\begin{figure}
\includegraphics[width=110mm]{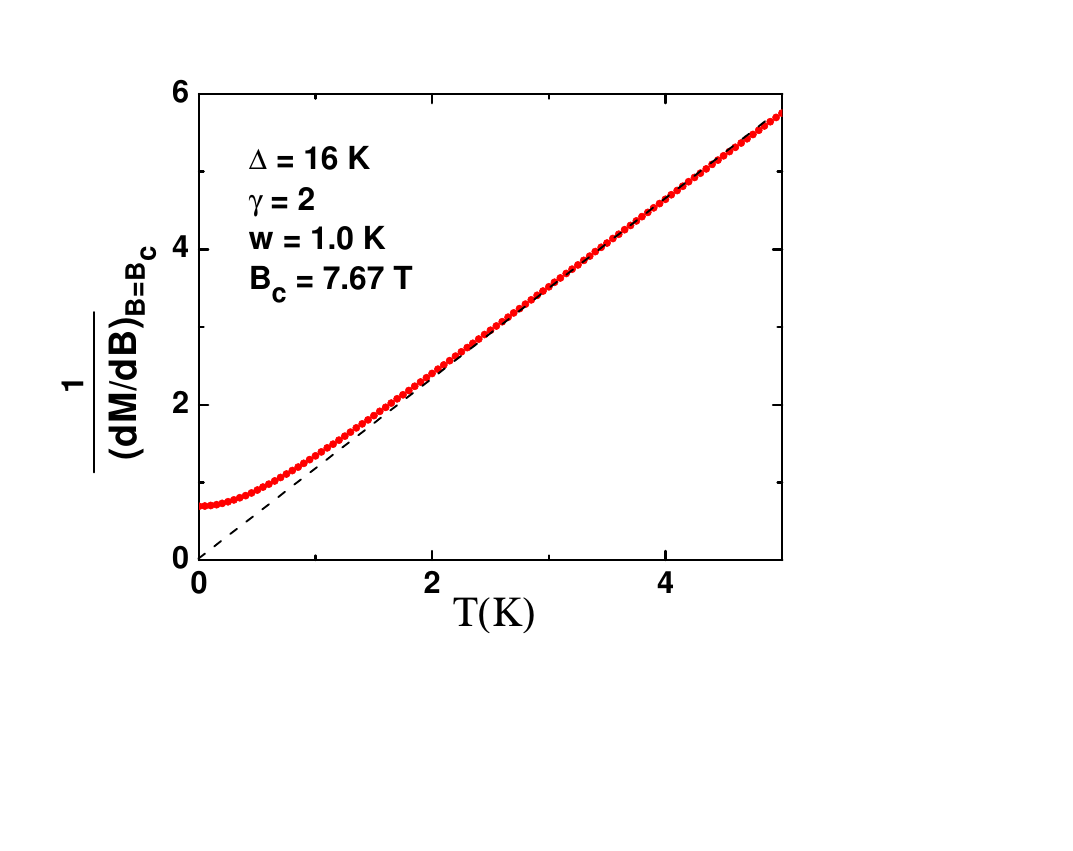}
\caption{\label{fig14} Shows the deviation from linearity with temperature of the inverse differential susceptibility at the critical field modeled by adding a level broadening of width $w$.  In this plot the parameter w has been chosen appropriate for CeRu$_2$Si$_2$. }
\end{figure}

While a level broadening scheme can account satifactorily for the high field behavior it fails quantitatively to account for the zero-field susceptibility.  With the value of $w$ that produces the correct saturation of $ (\partial m/ \partial b)^{-1}$ a non-zero value of $\chi_1(0)$ does indeed result but is too small compared with experiments.  It is possible to rectify this situation i.e. preserve the small value of w needed to explain the critical field behavior yet obtain a large $\chi_1(0)$ if the ground state is postulated to be a doublet with a concurrent lifting of its degeneracy as shown by the dotted line inside the hatched region in Fig.~6a.  It appears that to obtain such a doublet one has to go beyond effective spin 1 models (see for example \cite{ShivaramJCMP2017}).  

\begin{figure}
\includegraphics[width=110mm]{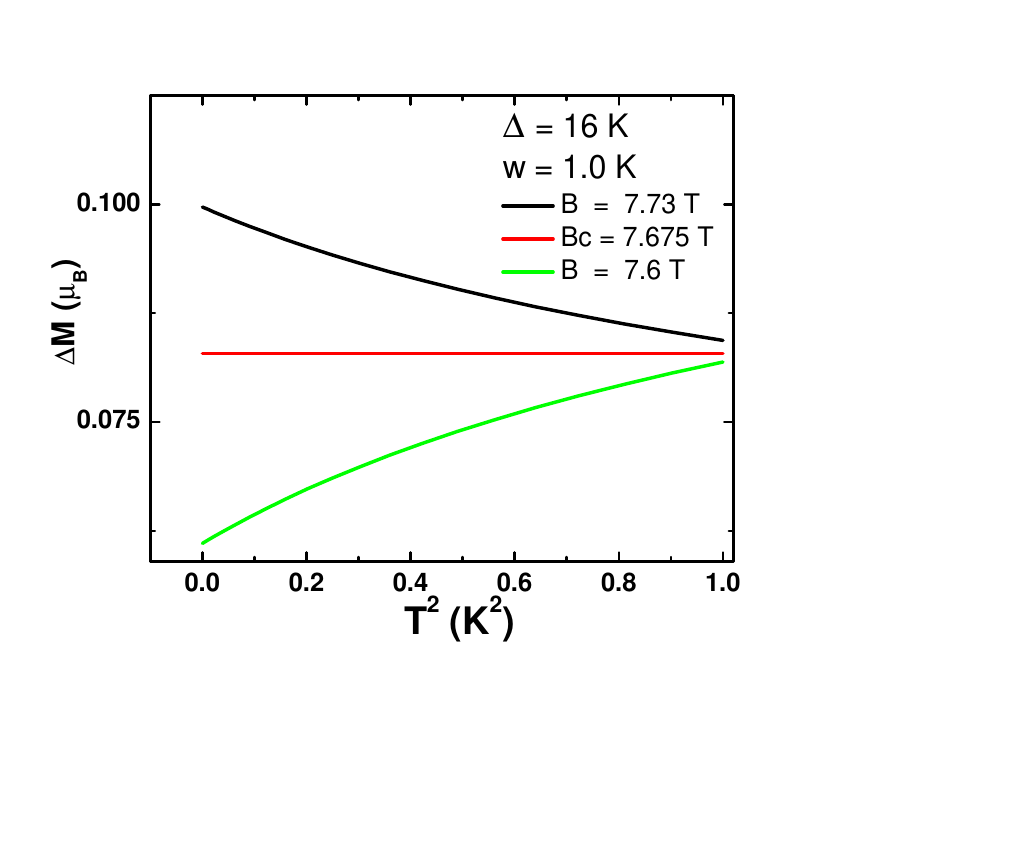}
\caption{\label{fig15} Shows the behavior of the magnetization in the very low temperature (T$<$1K) region obtained in the model with a level broadening $w$ included.  In this low temperature limit the magnetization is observed to obey a $T^2$ dependence, a behavior readily obtained from the Maxwell relation $ \partial M/ \partial T = \partial S/\partial B$ since $S \sim T$ for a Fermi system.  In this plot again the parameters have been chosen appropriate for CeRu$_2$Si$_2$. To accommodate all three curves on the same plot an arbitrary and different offset is applied on the vertical scale (hence $\Delta$M).}
\end{figure}

As an alternate to the level broadening scheme interactions can be taken into account by introducing an off-diagonal matrix element $v$ that mixes the gound-state singlet with the doublet of the Hamiltonian (2).  As shown in Section IIc, a magnetic field tilted at an angle $\theta$ from z leads to an avoided crossing at $b=1$, and hence to a broadened metamagnetic transition.  The same obtains for any matrix element $v$, leading to the eigennvalue equation

\begin{equation} 
\lambda ^3-2 \lambda^2+{\lambda}\left(1-b^2-2v^2\right)+2v^2 =0 , \label{eq:Eqwithv}
\end{equation}

This is the same as Eq.~(14) with $b^2 \rightarrow b^2 + 2v^2$ and $b_{x}^2 \rightarrow 2v^2$ (which implies $b_{z}^2 \rightarrow b^2$). The same replacement can be made in all the remaining formulas of Section IIc. In particular, Eq.~(\ref{eq:AvoidedCrossing}) becomes the standard two-level formula
$$\lambda_{1,2} \approx \frac{1}{2}\left(1-b_z \pm \sqrt{(1-b_z)^2+4v^2}\right).$$ \\

The anticrossing illustrated in fig.7 can also be attributed to a certain value of 'v' rather than the tilt as specified.

\section{Discussion}Strongly correlated heavy fermions come in a variety of crystal structures and weakly magnetic ground states. The many body interactions and crystal structures lead to a specific energy level sequence, which can be written down as a spin Hamiltonian, with its own characteristic anisotropy. The latter can then be studied for its response to a magnetic field in a given direction with respect to the crystalline axes. The nonlinear part of this response has also been calculated within the context of the Anderson Model though with difficulty \cite{BauerPRB2006}. Such calculations are material specific. Yet the measured properties are often universal. They involve a small number of energy scales. In that sense, the attempt in this paper is the proverbial first step in search of a minimal spin Hamiltonian. We have shown, to lowest order, that the phenomena of metamagnetism can be described in terms of an S = 1 pseudospin with one energy scale $–$ that of a singlet ground state separated from a doublet. The lower state of the doublet, under a magnetic field crosses the singlet and at that point we have a large magnetization response thus defining the critical field. These are the features of a typical hard axis spin Hamiltonian.

To summarize, in a minimal model, with a spin only Hamiltonian for metamagnetism where the critical field is shown to be related to the peak temperatures in nonlinear susceptibilities, the model captures many of the observed features in the thermodynamic properties and the sound velocity.  The model obviously excludes any treatment of transport properties.  Instead it assumes that magnetic properties including magnetoelastic effects, can be effectively described independently of transport.  Magnetostriction is simply accounted for by the parameter $\beta =\frac {dln\Delta}{dV}$.  In this paper only the uniform dilation and the corresponding sound velocity are considered.  But the extension to arbitrary strain is straightforward.  

We have also briefly discussed the inclusion of other degrees of freedom, to the extent that they modify the results of the simple one spin localized model.  Interactions with spins in other sites are treated in the mean field approximation.  They can significantly shift the critical field for metamagnetism and even lead to a phase transition that is a mix of meta and ferro (or anti-ferro) magnetism.  We plan to examine this mixed transition in future work.  All other effects are accounted for, phenomenologically by a parameter $w$.  Thus in the final analysis the full model involves several parameters, in addition to $\Delta$, and $\gamma$ that are absorbed by rescaling into "universal" plots.  There are $\beta$, $\lambda$ (for the mean field strength) and $w$ and others that we have mentioned and it is possible that additional parameters (for instance an anisotropy of the g-factor) will be required to fully describe all the complexities of heavy fermion materials.  Also, in our presentation we have considered incorporating into the equations one parameter at a time.  In the end it will be left to the experimentalists to determine when all presented aspects are considered the extent to which this model falls short in describing their data.  Whether useful predictions to aid in the discovery of future heavy fermion materials from this model remains to be seen.

%

\appendix
\section{Negative $\Delta$}
    If $\Delta <0$ in the minimal Hamiltonian $\mathcal{H}$ of Eq.~(\ref{eq:Hmin}), the spin likes to point in the $z$ direction. The eigenvalues are as shown in Fig. 6a, turned upside down, with the understanding that now $b=\gamma B/|\Delta |$. 
The equations of Section III remain valid for negative $\Delta$; in the scaled variables, simplly change the sign of $b$ and $\tau$, and also of $\chi $ (snce it is defined as $\partial m/\partial b$.\\
Noting that $-S_{z}^2 $ is equivalent to $S_x^2 + S_y^2 $, we expect that negative      $\Delta$ will give "weak metamagnetism" when $\bm B$ is in the $xy$-plane.
This is now the interesting case, and we discuss it below, leaving aside the
simple near-Curie behavior for $\bm B$ along $z$ (and also the gradual change with angle from z to x).  \\
\subsection{B along x}

\begin{figure}
\includegraphics[width=110mm]{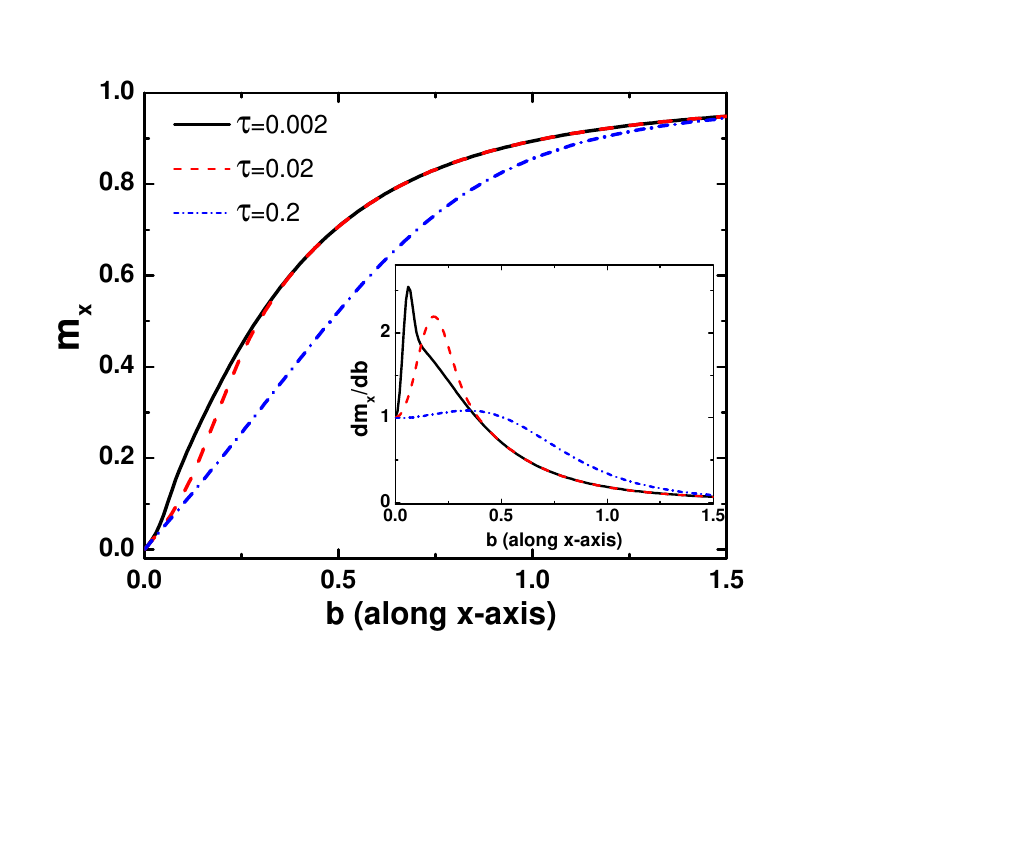}
\caption{\label{fig16}:  Shows the magnetization isotherms obtained in the minimal model when $\Delta$ is negative.  A weak metamagnetism is seen which becomes clear when the derivative is considered - see inset.}.
\end{figure}

The dimensionless magnetization, $m$, is given by Eq.~(\ref{eq:mforBx}), but with $e^{1/2\tau}$ replacing $e^{-1/2\tau}$ in the denominator. As a consequence, the  isotherms are depressed, compared to those of Fig.~4, and at low $T$ have an inflection point, as seen in Fig.~13.   In this respect, they resemble those of Fig.~1, for $\Delta >0$ in the metamagnetic direction $z$ . However, there is no critical point at $b=1$, no curve crossing, and the inflection point already disappears at $\tau=0.2$.  \\

\end{document}